# Performance Evaluation of Shared Hosting Security Methods

*Seyed Ali Mirheidari, Sajjad Arshad, Saeidreza Khoshkdahan*

Computer Engineering Department, Sharif University of Technology, International Campus, Iran
Electrical and Computer Engineering Department, Shahid Beheshti University, General Campus, Iran
Sabzfaam Information Technology Corporation, Iran

mirheidari@kish.sharif.edu, s.arshad@mail.sbu.ac.ir, khoshkdahan@sabzfaam.ir

*Abstract*—Shared hosting is a kind of web hosting in which multiple websites reside on one webserver. It is cost-effective and makes the administration easier for websites' owners. However, shared hosting has some performance and security issues. In default shared hosting configuration, all websites' scripts are executed under the webserver's user account regardless of their owners. Therefore, a website is able to access other websites' resources. This security problem arises from lack of proper isolation between different websites hosted on the same webserver. In this survey, we have examined different methods for handling mentioned security issue. Also we evaluated the performance of mentioned methods. Finally, we evaluated performance of these methods with various configurations.

## I. INTRODUCTION

Formerly, many people published their websites on their own dedicated servers. But, with the increase of hardware power, it is possible to host thousands of websites simultaneously on only one server. As a result, the customers do not have to prepare an individual server to publish their websites and there is no need to have lots of knowledge and experience for administration. This kind of web hosting is called shared hosting. Nevertheless, shared hosting has security problems which occur because of the webserver's privilege during execution. Webserver must have access to every websites' files in order to serve the websites and their scripts have to be run with webserver privilege. Therefore, websites' owners are able to access other websites' files. In particular, this security flaw is serious for a hosting service where many websites are hosted on a server. Setting the proper permissions is a trivial task because the webserver software runs as a particular user who needs to have access to every file. The permission and its related issues has involved administrators for many years and the world has witnessed increasing number of web sites defacement [1]. Malicious users can exploit the ability of the webserver process to have access to server resources as well as websites' files. Exploiting this vulnerability is usually done by some malicious scripts which have been developed to cover this need. According to Zone-H[1] site, a noticeable number of defacements are released in only one site [2,3]. This statistic would be even more bothering when many shared hosts are completely defaced because of the above mentioned security problem. Several methods have been proposed to decrease the risk of this security problem which in many cases were inefficient or in some other cases impose performance limitations. The main goal of this survey is to introduce and examine various methods and express their weaknesses and strengths. It also pays much more attention to their performance and ability to solve the security problem.

According to the newest monthly survey in October 2011 from Netcraft [4], Apache[2] webserver has the first rank of use 64.67%, and Microsoft IIS comes second with only 15.66%. As Apache is widely used, in this survey we consider the security problem and methods in this popular webserver. Since the most methods have been released for POSIX[3] operating systems (especially specific distribution of Linux), our survey is mainly focused on Linux. We use PHP[4] along with Apache as a web application because of higher popularity, usability and reliability on these platforms; in other words, we deal with LAMP[5] servers.

The remainder of this paper is outlined as follows. Section II presents the shared hosting architecture and security problem. In section III, we discuss the various methods for solving the security problem. Section IV evaluates the performance and scalability of the methods. Finally, the paper is concluded in section V.

## II. SHARED HOSTING

We can utilize server resources by sharing them among different websites. There are two methods of achieving this goal: 1) Using a dedicated webserver for each website which is installed on a virtual machine or 2) Using a single webserver to host many websites also known as virtual websites. The first method isolates webservers and provides good security, but does not utilize resources well. The second method uses resources efficiently and is flexible. The following sections will illustrate virtual website method.

---

[1] http://www.zone-h.org

[2] http://www.apache.org

[3] Portable Operating System Interface for Unix

[4] http://www.php.net

[5] Linux Apache MySQL PHP

*A. Architecture*

In shared hosting, there are many websites on a single webserver. Each website has an owner and each customer has a FTP account which can upload new files for the website and the uploaded files are owned by customer's user account. Apache webserver runs as a specific user (www-data) and handles all HTTP requests for all websites. Apache needs to be able to read the files on each website. Figure 1 shows the necessary permissions for websites' directories.

```
rwx --x ---    web1:www-data    /home/web1/
rwx r-x ---    web1:www-data    /home/web1/public_html
rwx --x ---    web2:www-data    /home/web2/
rwx r-x ---    web2:www-data    /home/web2/public_html
```

Figure 1. Setting permission for websites' directories

In some Content Management System (CMS) applications, Apache needs the write access to websites' directories, too.

*B. Security Problem*

In shared host servers, Apache has access to every website's files and Apache user has read/execute permissions on every website's directory. Therefore, if an attacker breaks into one of the websites on the webserver, he can have access to other websites' directory hosted on the same webserver. Also, if an attacker registers a website on the server as a legitimate customer, he can write some scripts to access other websites on that server. Some common security problems are file system browsing, write access, tampering, session poisoning and session snooping [5].

*C. Challenge*

Many methods have been introduced for above mentioned security problem in different layers. But the question is which of them can solve the security problem with the best performance? Stuart Herbert [6] in his weblogs calls it "The Challenge with Securing Shared Hosting" and he tried to examine the performance of variety of methods.

III. METHODS AND LIMITATIONS

There are some methods to secure shared web hosting. In this section, we will examine the common methods developed for apache and explain their main idea to solve the security problem.

*A. PHP Methods*

PHP developers are trying to overcome a security problem by proposing two methods in PHP layer which are Safe_Mode and Open_Basedir. Both of these methods have some limitations. In other words, PHP is not the right place to address the security problem.

*1) Safe_Mode*

In Safe_Mode [7], PHP examines the access of running PHP scripts to the files based on their owners. PHP checks the owner of those files and if the owner of the file is not the same as the owner of the running script, PHP will not allow that access. But, Safe_Mode has a few limitations. There are some applications that upload files to server. The owner of those files will be Apache user, not the script owner user. Afterward, those files cannot be accessed by the PHP scripts.

*2) Open_Basedir*

In Open_Basedir [8], PHP determines the directory which each user is allowed to access. PHP examines the file access of running PHP scripts and do not allow access to files outside that directory.

*B. Apache Module Methods*

If we examine the security problem closely, we can find out that the main cause of the problem is how we run the Apache server. As you can see, Apache is executed by a unique user who can have access to every websites' files. A new idea is that Apache serves each website by its owner user account. In other words, Apache runs each script with its owner permissions. There are two common methods, suEXEC and suPHP, which use this idea and have been developed as an Apache module. In the following sections, we examine these methods in details.

*1) suEXEC*

The suEXEC [9] is composed of a wrapper binary file and an Apache module. When an HTTP request arrives, Apache runs the wrapper and gives the script name and user/group ID under which the script has to be executed to the wrapper. The suEXEC can only be used with CGI [10] or FastCGI [11] programs. In order to use suEXEC, we need a unique CGI or FastCGI binary file for each website. The user/group ID of the owner must be the owner of website. We should update these binary files when the new version of PHP is released. Also, if we use PHP in CGI or FastCGI mode, we cannot employ HTTP authentication feature. Using suEXEC with CGI has very low performance in a way that Corentin Chary has named it as a performance killer [12].

*2) suPHP*

Same as suEXEC, suPHP [13] runs PHP scripts with the specified user/group ID. The suPHP has an Apache module and a setuid-root binary file. Unlike suEXEC, by using suPHP, there is no need to have a unique CGI or FastCGI binary file for each website. Same as suEXEC, suPHP suffers from low performance.

## C. Apache MPM Methods

After Apache 2.0 has been released, various MPM [14] methods have been introduced to solve the shared hosting security problem. We examine these methods with greater details in the following sections.

### 1) Perchild and Metux MPMs

Perchild MPM [15] is the first MPM method to shared hosting security problem. The approach of Perchild is to run an Apache process for each website with user/group ID of the website's owner and each Apache process creates several threads to serve the requests for that website. For some reasons, this MPM has not been implemented but Enrico Weiglet implemented Metux MPM [16] based on Perchild MPM approach. Nevertheless, Metux MPM is not appropriate for PHP websites because of non-thread-safety nature of PHP.

### 2) Peruser MPM

Considering the fact that Metux MPM is not appropriate for PHP, Sean Gabriel Heacock introduced Peruser MPM [17]. Peruser MPM uses processes instead of threads to handle requests. Peruser MPM runs a control Apache process as root privilege. The control process creates several multiplexer processes with Apache user privilege. The multiplexer process listens to port 80 and accepts incoming connections and reads the request to check from which website it is. Then, it passes the connection to related worker process to handle it. The worker processes run under the user/group ID of respective websites' owners. The control process always maintains a pool of idle worker processes to enhance the performance and forks off new worker processes if there are no idle processes to handle new requests. One important shortcoming of Peruser MPM is excessive use of server resources.

### 3) ITK MPM

In order to reduce the shortcoming of Peruser MPM, Steinar Gunderson introduced ITK MPM [18]. ITK MPM creates a managing Apache process with root privilege. The managing process spawns several listener Apache processes with root privilege. The listener process listens to port 80 and reads new request to determine which website it is. Then, it creates a new Apache handler process with user/group ID of website's owner to serve the request. But, the main difference of ITK MPM with Peruser MPM is that after the request has been completed, the handler Apache process is terminated. In other words, ITK MPM doesn't maintain a pool of idle handler processes for serving the requests.

## IV. EVALUATION

The methods declared in previous sections can join some methods of execution to have practical solutions. Table I shows 16 possible solutions to secure shared host servers. In these solutions, PHP refers to PHP module in Apache webserver.

TABLE I. APACHE SOLUTIONS

| # | Name | Abbreviation |
|---|------|--------------|
| 1 | suEXEC-Prefork-CGI | suE-Pf-C |
| 2 | suEXEC-Prefork-CGId | suE-Pf-Cd |
| 3 | suEXEC-Prefork-FastCGI | suE-Pf-FC |
| 4 | suEXEC-Worker-CGI | suE-W-C |
| 5 | suEXEC-Worker-CGId | suE-W-Cd |
| 6 | suEXEC-Worker-FastCGI | suE-W-FC |
| 7 | suEXEC-Event-CGI | suE-E-C |
| 8 | suEXEC-Event-CGId | suE-E-Cd |
| 9 | suEXEC-Event-FastCGI | suE-E-FC |
| 10 | suPHP-Prefork | suP-Pf |
| 11 | suPHP-Worker | suP-W |
| 12 | suPHP-Event | suP-E |
| 13 | Peruser-PHP | Pu-PH |
| 14 | Peruser-CGI | Pu-C |
| 15 | ITK-PHP | I-PH |
| 16 | ITK-CGI | I-C |

The most important goal of this survey is to extract the fastest and the most secure solution among the all. In order to evaluate the performance of security solutions, we configured each solution with various number of virtual hosts and benchmarked the average response time, throughput, CPU usage, memory usage and number of processes. The hardware and software configuration for the performance evaluation are listed in Table II and Table III. To measure the performance of the solutions, we used Httperf benchmark [19] version 0.9 which send request to specific PHP script that calls *phpinfo()*, which displays the system information of the PHP language processor. The traffic generated by this function is about 47.9 KB per request. In order to have accurate results, we repeated the test 10 times and averaged the results. Whereas, the evaluation results of the solutions using Event MPM [20] and Worker MPM [21] is similar, we only brought the results of solutions using Worker MPM and Prefork MPM [22].

TABLE II. CLIENT/SERVER HARDWARE CONFIGURATION

| CPU | Intel Pentium 4, 3 GHz |
|-----|------------------------|
| Memory | 1024 MB |
| OS | Ubutnu 10.04 (Linux 2.6.32) |

TABLE III. SOFTWARE CONFIGURATION

| Apache (Prefork MPM, Worker MPM, Event MPM, CGI, CGId, suEXEC) | 2.2.16 |
|---|---|
| PHP | 5.3.6 |
| FastCGI | 2.3.6 |
| suPHP | 0.7.1 |
| Peruser MPM | 0.4.0rc2 |
| ITK MPM | 2.2.16 |

## A. Performance

We evaluated the performance of above mentioned solutions with different request frequencies to determine their effectiveness. To evaluate the performance, we configured each solution with one virtual host and sent concurrent requests to the server. We measured the throughput and response time of each solution with different numbers of concurrent requests (1, 10, 20, 30, 40, 50, 60, 70, 80, 90 and 100). We calculated the throughput as the average numbers of requests which webserver can handle in one second and response time as the average time of processing a request. The throughput and response time of various solutions are shown in Figure 2. As you can see, the throughput of solutions using FastCGI does not degrade up to 90 requests per second. Also, the response time of these solutions are almost near zero. After these solutions, the Peruser-PHP solution has the highest throughput and its throughput degrades after 70 requests per second. The average response time of this solution is very low and is about 2 seconds. The remaining solutions behave similarly after 20 requests per second and their throughputs are less than the 20 requests per second. Also, the average response time of these solutions is higher than 10 seconds after 20 requests per second. Therefore, the solutions using FastCGI and the Peruser-PHP solutions are appropriate for webservers which have high request frequency.

## B. Scalability

We evaluated the scalability feature of solutions with high number of virtual hosts to determine resource usage of each solution. To evaluate the scalability, we configured each solution with 100 virtual hosts and run 100 Httperf instances simultaneously. Then, each Httperf instance sent 100 requests to the corresponding virtual host sequentially and we measured the maximum and average CPU usage, memory usage and the number of created processes for each solution during the evaluation. Also, some solutions maintain a pool of idle processes after the completion of the evaluation. Therefore, we measured the number of idle processes which occupied the memory and the amount of memory that they use. In following sections, we analyze the evaluation results.

### 1) CPU Usage

The CPU usage of each solution is shown in Figure 3. As you can see, the difference between the highest and lowest average CPU usage is almost 8%. The solutions using suPHP have higher CPU usage on average. Because suPHP doesn't maintain a pool of idle processes, it has to create a new handler CGI process and it really affects the CPU usage. On the other hand, the solutions using FastCGI have lower CPU usage on average because of the fact that FastCGI maintains a pool of idle CGI processes to handle new requests so it does not need to create a new CGI process for each incoming request and terminate it after the completion.

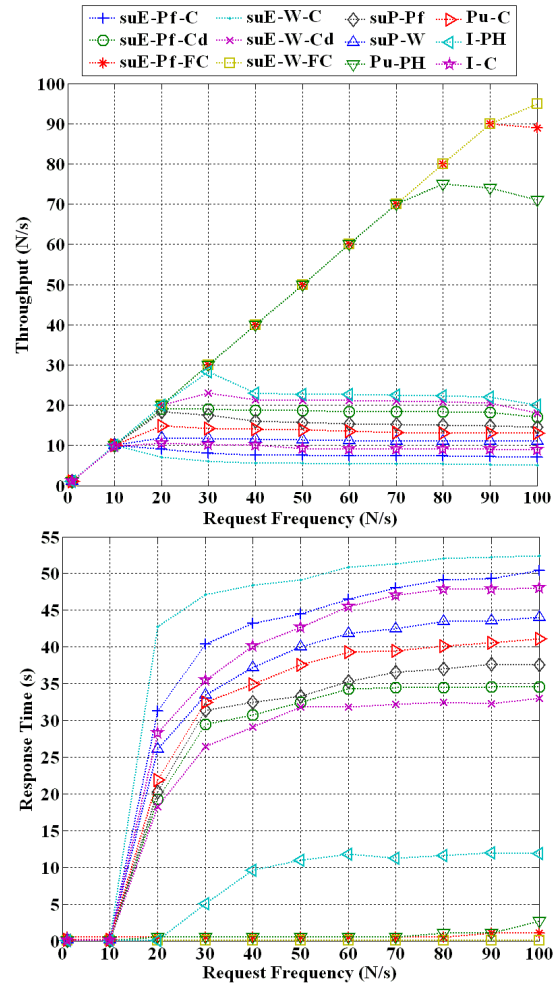

Figure 2. Throughput and Response Time

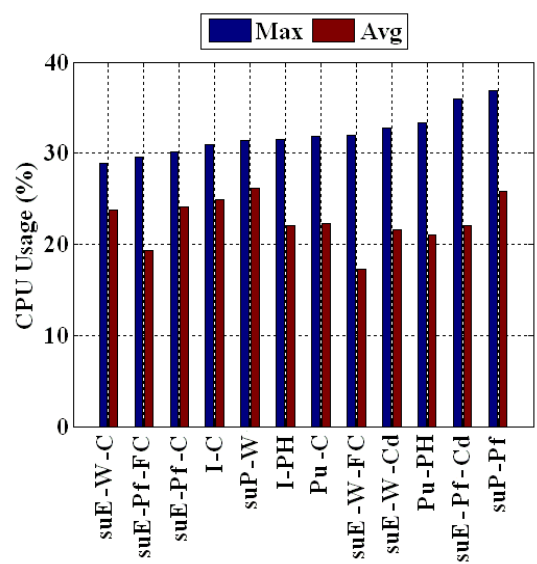

Figure 3. CPU Usage

*2) Memory Usage*

As shown in Figure 4, the difference between the highest and lowest average memory usage is about 90 MBs. The solutions using Peruser MPM or FastCGI have higher memory usage because they maintain a pool of idle workers for handling new incoming requests. Also, these solutions do not free memory and maintain their idle processes in memory. On the other hand, the solutions using CGId [23] have lower memory usage. The solution using ITK MPM have a low memory usage in comparison to the solutions using Peruser MPM and this is an important advantage.

*3) Number of Processes*

The number of processes for each solution is shown in Figure 5. As can be seen, the difference between the highest and lowest average number of processes is almost 305. The Peruser-CGI and ITK-CGI solutions have higher number of processes because for handling a request, they use two processes, an Apache process with website's user account privilege and a CGI process which process the request. On the other hand, the solutions using Worker MPM have lower number of processes because of the fact that the Worker MPM uses threads for handling requests. Finally, after the completion of the evaluation, the solutions using Peruser MPM or FastCGI, keep pool of idle processes in memory.

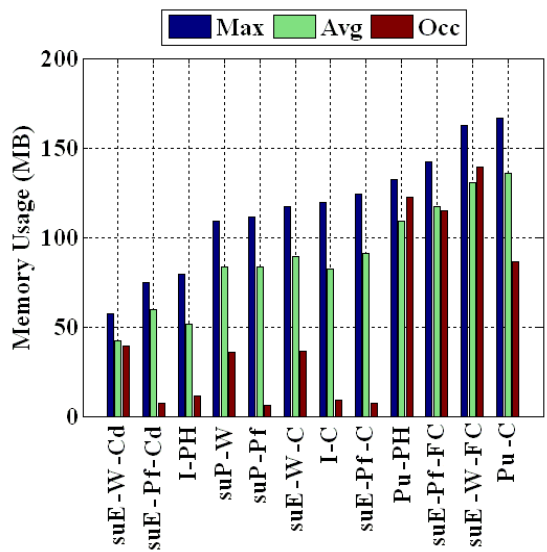

Figure 4. Memory Usage

C. *Security*

Although all mentioned methods can solve the security problem, they impose a security risk to the whole webserver. As stated previously, suEXEC and suPHP have a wrapper which executes with root privileges to create a CGI process with the corresponding website's owner identity. Therefore, there is a potential security risk that attackers can escalate their privileges to root user. To overcome the security risk of suPHP, Marco Prandini and et al [24] have presented a SELinux-based method. Also, Peruser and ITK MPMs have processes listening on port 80 with root privileges. This is risky because it leaves the risk of a security hole allowing someone to gain root access to the machine. Overall, we cannot eliminate this security risk under the Discretionary Access Control (DAC) model implemented on Linux and we have to exploit the features of Mandatory Access Control (MAC) systems.

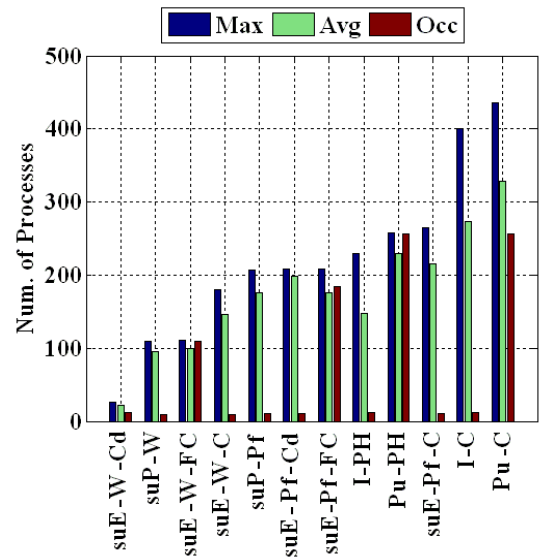

Figure 5. Number of Processes

D. *Comparison*

The results of comparing solutions from different points of view are shown in Table IV.

TABLE IV. COMPARISON OF SOLUTIONS

| Solution | Performance | Scalability | Security |
|---|---|---|---|
| suE-Pf-C | Very Low | High | High |
| suE-Pf-Cd | Low | High | High |
| suE-Pf-FC | Very High | Moderate | High |
| suE-W-C | Very Low | High | High |
| suE-W-Cd | Low | High | High |
| suE-W-FC | Very High | Moderate | High |
| suE-E-C | Very Low | High | High |
| suE-E-Cd | Low | High | High |
| suE-E-FC | Very High | Moderate | High |
| suP-Pf | Very Low | High | High |
| suP-W | Very Low | High | High |
| suP-E | Very Low | High | High |
| Pu-PH | High | Low | Moderate |
| Pu-C | Low | Low | Moderate |
| I-PH | Moderate | Very High | Moderate |
| I-C | Very Low | High | Moderate |

As you can see, none of the solutions behave very well in every aspect and we have to use each solution based on the shared hosting server's requirements. If we want a high performance shared hosting server, we have to employ the Peruser-PHP solution and the solutions using FaseCGI. For handling high number of websites, the solutions using CGI or CGId or ITK are appropriate. If the security aspect is vital, using the

MPM solutions are not recommended. Also, for reliability, you can use all of the solutions except the solutions using FastCGI. Finally, if the complexity of configuration and maintenance is important, it is better not to use Peruser MPM solutions and the solutions using FastCGI. As a whole, if we want to have a trade-off between different aspects and have a solution which has a relatively good behavior in all of them, the ITK-PHP solution is a good choice.

V. CONCLUSION

Although shared hosting is a great solution for sharing server's resources between several websites, but it has some drawbacks and security weaknesses which should be minimized in order to derive the benefits it offers. The security problem arises because Apache has access to every website's files and serves them with the same identity. In order to overcome the security problem, several methods have been released. We enumerated 16 practical solutions and evaluated their performance and scalability. Considering the comparison given in Table IV, the ITK-PHP solution behaves relatively well in all of the aspects and it is a good choice for shared host servers. Although, all of the proposed methods solve the security problem, they impose the security risk of root privilege escalation by intruder. Generally, we cannot resolve the security risk only by using DAC model and we have to take the advantage of MAC systems.

REFERENCES

[1] (2010) Zone-H: Defacements Statistics 2008 - 2009 - 2010*. [Online]. http://zone-h.com/news/id/4735
[2] (2011, January) Zone-H: Defacements Statistics 2010: Almost 1,5 million websites defaced, what's happening? [Online]. http://zone-h.com/news/id/4737
[3] (2011) Zone-H: Yearly & Mounthly & Daily Attacks statistics. [Online]. http://www.zone-h.org/stats/ymd
[4] Netcraft: November 2011 Web Server Survey. [Online]. http://news.netcraft.com/archives/2011/11/07/november-2011-web-server-survey.html
[5] N. Nikiforakis, W. Joosen, and M. Johns, "Abusing Locality in Shared Web Hosting," in *European Workshop on System Security*, Salzburg, Austria, 2011.
[6] H. Stuart. (2007) The Challenge with Securing Shared Hosting. [Online]. http://blog.stuartherbert.com/php/2007/11/21/the-challenge-with-securing-shared-hosting/
[7] PHP: Safe_Mode. [Online]. http://php.net/manual/en/features.safe-mode.php
[8] PHP: Open_basedir. [Online]. http://www.php.net/manual/en/ini.core.php#ini.open-basedir
[9] Apache: suEXEC. [Online]. http://httpd.apache.org/docs/2.0/suexec.html
[10] Apache: CGI. [Online]. http://httpd.apache.org/docs/2.2/howto/cgi.html
[11] Apache: FastCGI. [Online]. http://httpd.apache.org/mod_fcgid/
[12] C. Chary and C. Khamly. Securing A Shared Web Server. [Online]. http://xf.iksaif.net/papers/securing-a-shared-web-server.pdf
[13] suPHP. [Online]. http://www.suphp.org/Home.html
[14] Apache: Multi-Processing Module (MPM). [Online]. http://httpd.apache.org/docs/2.0/mpm.html
[15] Apache: Perchild MPM. [Online]. http://httpd.apache.org/docs/2.0/mod/perchild.html
[16] Metux MPM. [Online]. http://www.sannes.org/metuxmpm/
[17] Peruser MPM. [Online]. http://www.peruser.org
[18] S. H. Gunderson. The Apache 2 ITK MPM. [Online]. http://mpm-itk.sesse.net/
[19] D. Mosberge and T. Jin, "Httperf: A Tool for Measuring Web Server Performance," in *1st Workshop on Internet Server Performance*, 1998.
[20] Apache: Event MPM. [Online]. http://httpd.apache.org/docs/2.2/mod/event.html
[21] Apache: Worker MPM. [Online]. http://httpd.apache.org/docs/2.0/mod/worker.html
[22] Apache: Prefork MPM. [Online]. http://httpd.apache.org/docs/2.0/mod/prefork.html
[23] Apache: CGId. [Online]. http://httpd.apache.org/docs/2.0/mod/mod_cgid.html
[24] M. Prandini, E. Faldella, and R. Laschi, "Mandatory Access Control applications to web hosting," in *Second European conference on computer network defence, Pontypridd (Cardiff), Wales*, London, 2006.